\documentstyle[12pt,aps]{revtex}
\tightenlines
\begin{document}
\centerline{\bf Transfer matrices and lattice fermions at finite density}
\medskip
\centerline{Michael Creutz}
\centerline{creutz@bnl.gov}
\medskip
\centerline{Brookhaven National Laboratory}
\centerline{Upton, NY 11973, USA}
\bigskip
\begin{abstract}
I discuss the connection between the Hamiltonian and path integral
approaches for fermionic fields.  I show how the temporal Wilson
projection operators appear naturally in a lattice action.  I also
carefully treat the insertion of a chemical potential term.
\end{abstract}
\bigskip
It is a pleasure to contribute to this volume honoring Kurt Haller.
Kurt and I have long shared strong interests in the connection between
Hamiltonian and Lagrangian formulations of field theory.  We have both
extensively investigated the ``temporal gauge'' with hopes of gaining
insight into non-perturbative phenomena in gauge
theories\cite{temporalgauge}.  Topological tunnelling issues in the
Hamiltonian formalism have been entwined with much of this activity.
While Kurt has emphasized the continuum formulation, for me many of
these ideas have become increasingly involved with lattice issues.
Actually my first paper on the temporal gauge concerned the
lattice\cite{transfer1}, showing the connection between the Lagrangian
formulation of Wilson and the Hamiltonian formulation of Kogut and
Susskind.

Here I expand on some of these old relations between the Hilbert space
approach and path integrals, with an emphasis on the complications
arising with anticommuting fields.  My initial paper on the transfer
matrix concentrated on gauge fields and was rather terse with
fermions\cite{transfer1}.  Later I returned to these issues, pointing
out how the Wilson projection operator formalism\cite{wilsonf}
naturally arises\cite{transfer2}.  Here I reconstruct that argument in
a more pedagogical manner, adding comments on how a chemical
potential, representing a non-zero baryon density, most naturally
appears in a fermionic path integral.

Anticommutation is at the heart of fermionic behavior.  This is true
in both Hamiltonian operator formalisms and Lagrangian path integral
approaches, but in rather complementary ways.  If an operator
$a^\dagger$ creates a fermion in some normalized state on the lattice
or the continuum, it satisfies the basic relation
\begin{equation}
[a,a^\dagger]_+\equiv a a^\dagger + a^\dagger a =1.
\end{equation}
This contrasts sharply with the fields in a path integral, which all
anticommute 
\begin{equation}
[\chi,\chi^\dagger]_+=0.
\end{equation}
The connection between the Hilbert space approach and the path
integral appear through the transfer matrix formalism.  For bosonic
fields this is straightforward, but for fermions certain subtlies
arise, related to the so called ``doubling problem.''

To be more precise, consider a single fermion state created by the
operator $a^\dagger$, and an antiparticle state created by another
operator $b^\dagger$.  For an extremely simple model, consider the
Hamiltonian
\begin{equation}
H=m (a^\dagger a+b^\dagger b) + \mu (a^\dagger a-b^\dagger b).
\end{equation}
Here $m$ can be thought of as a ``mass'' for the particle, and $\mu$
represents a chemical potential.  What I want is an exact path
integral expression for the partition function
\begin{equation}
Z={\rm Tr} e^{-\beta H}.
\end{equation}
Of course, since my Hilbert space has only four states, this is
trivial to work out: $ Z=1+e^{m+\mu}+e^{m-\mu}+e^{2m} $. However, I want
this in a form that easily generalizes to many variables, makes the
connection with the Wilson projection operator clear, and illustrates
how the chemical potential is properly inserted into a path integral.

A path integral for fermions uses Grassmann variables\cite{grassmann}.
I introduce a pair of such, $\chi$ and $\chi^\dagger$, which will be
connected to the operator pair $a$ and $a^\dagger$, and another pair,
$\xi$ and $\xi^\dagger$, for $b$, $b^\dagger$.  All the Grassmann
variables anticommute.  Integration over any of them is determined by
the simple formulas
\begin{equation}
\int d\chi \ 1 = 0\ ; \qquad \int d\chi\ \chi = 1.
\end{equation}
This determines my phase conventions.  For notational simplicity I
join individual Grassmann variables into spinors
\begin{equation}\matrix{
\psi = \pmatrix{\chi\cr\xi^\dagger\cr};&
\psi^\dagger = \pmatrix{\chi^\dagger&\xi\cr}.\cr
}
\end{equation}
To make things appear still more familiar, introduce a ``Dirac
matrix''
\begin{equation}
\gamma_0=\pmatrix{1&0\cr 0 & -1\cr}
\end{equation}
and the usual
\begin{equation}
\overline\psi=\psi^\dagger\gamma_0.
\end{equation}
Then we have
\begin{equation}
\overline \psi \psi = \chi^\dagger \chi + \xi^\dagger \xi.
\end{equation}
The temporal Wilson projection\cite{wilsonf} operators
\begin{equation}
P_\pm={1\over2}(1\pm\gamma_0)
\end{equation}
arise when one considers the fields at
two different locations
\begin{equation}
\chi_i^\dagger  \chi_j +\xi_i^\dagger  \xi_j =
\overline \psi_i P_+\psi_j+\overline \psi_j P_-\psi_i.
\label{equ:projection}
\end{equation}
The indices $i$ and $j$ will soon label the ends of a temporal hopping
term; this formula is the basic transfer matrix justification for the 
projection operator formalism.

For a moment I ignore the antiparticles and consider some general
operator $f(a,a^\dagger)$ in my Hilbert space.  How is this
related to an integration in Grassmann space?  To proceed I need a
convention for ordering the operators in $f$.  I adopt the usual
normal ordering definition with the notation $:f(a,a^\dagger):$
meaning that creation operators are placed to the left of destruction
operators, with a minus sign inserted for each exchange.  In this case
a rather simple formula gives the trace of the operator as a Grassmann
integration
\begin{equation}
{\rm Tr}\ :f(a,a^\dagger):\ = \int d\chi d\chi^\dagger 
e^{2\chi^\dagger\chi} f(\chi,\chi^\dagger).
\end{equation}
To verify, just check that all elements of the complete set of
operators $\{1,a,a^\dagger,a^\dagger a\}$ work.  However,
this formula is actually much more general; with a set of Grassmann
variables $\{\chi,\chi^\dagger\}$, one pair for each fermion state,
this immediately generalizes to the trace of any normal ordered
operator acting in a many fermion Hilbert space.

What about a product of several normal ordered operators?  This leads
to the introduction of multiple sets of Grassmann variables and the
general formula \cite{transfer2,thomaz,luscher}
\begin{eqnarray}
&&{\rm Tr}\ :f_1(a^\dagger,a):\ :f_2(a^\dagger,a): \ldots :f_n(a^\dagger,a):
\ =\cr
&&\int d\chi_1\ d\chi_1^*\ldots d\chi_n\ d\chi_n^*\cr
&&e^{\chi_1^*(\chi_1+\chi_n)} e^{\chi_2^*(\chi_2-\chi_1)} 
 \ldots e^{\chi_n^*(\chi_n-\chi_{n-1})} \cr 
&&f_1(\chi_1^*,\chi_1)f_2(\chi_2^*,\chi_2)
 \ldots f_n(\chi_n^*,\chi_n).
\label{equ:main}
\end{eqnarray}
The positive sign on $\chi_n$ in the first exponential factor
indicates the natural occurance of antiperiodic boundary conditions.
With just one factor, this formula reduces to the previous relation.
Note how the ``time derivative'' terms are ``one sided;'' this is how
doubling is eluded.

This exact relationship provides the starting place for converting our
partition function into a path integral.  The simplicity of the
Hamiltonian allows this to be done exactly at every stage.  First I
break ``time'' into a number $N$ of ``slices''
\begin{equation}
Z={\rm Tr} \left( e^{-\beta H/N}\right)^N. 
\end{equation}  
Now I need normal ordered factors for the above formula.  For this I
use
\begin{equation}
e^{\alpha a^\dagger a} = 1+(e^\alpha-1) a^\dagger a
=\ :e^{(e^\alpha-1) a^\dagger a }:\ ,
\end{equation}
true for arbitrary parameter $\alpha$.  Combining the particles and
antiparticles into one matrix equation gives
\begin{equation}
\exp((\alpha+\rho) a^\dagger a+(\alpha-\rho) b^\dagger b) =
\ : \exp\left(\pmatrix{a^\dagger & b^\dagger\cr}
\left(e^{\alpha+\rho\gamma_0}-1\right)
\pmatrix{a\cr b\cr}\right):\ ,
\end{equation}
The $-1$ in this exponent will cancel the diagonal terms in the
exponentials of Eq.~(\ref{equ:main}).

This is all the machinery I need to write
\begin{equation}
Z=\int (d\psi d\psi^\dagger) e^{S}
\end{equation}
where
\begin{equation}
S=\sum_{i=1}^n \overline\psi_n e^{-\beta m/N}e^{-\beta \mu \gamma_0 /N}
\psi_n
-\overline\psi_n P_+ \psi_{n-1}
-\overline\psi_{n-1} P_ - \psi_{n}.
\label{equ:mvm}
\end{equation}
Note how the Wilson projection factors of $P_\pm$ automatically appear
via Eq.~(\ref{equ:projection}) to handle the reverse convention of
$\chi$ versus $\xi$ in our field $\psi$.  The projection operator
formalism is a natural consequence of an exact transfer matrix.

The chemical potential appears simply as an inserted factor of
$e^{-\beta \mu \gamma_0 /N}$ in the ``mass'' term.  This is not quite
in the conventional form \cite{chempot} since the chemical potential
piece is temporally diagonal.  However this is actually only a
convention, as the factor can be moved to temporal links with a change
of variable
\begin{equation}
\Psi \equiv e^{-\beta\mu\gamma_0\over2N} \psi.
\end{equation}
With this substitution, we have 
\begin{equation}
S=\sum_{i=1}^n \overline\Psi_n e^{-\beta m/N}
\Psi_n
-\overline\Psi_n P_+ e^{\beta\mu/N} \Psi_{n-1}
-\overline\Psi_{n-1} P_- e^{-\beta\mu/N}\Psi_{n}.
\label{equ:onlinks}
\end{equation}
If $\mu$ were imaginary, this would be precisely the form of a $U(1)$
gauge field on the timelike bonds.

In this discussion I have ignored spatial hoppings.  Terms in the action
of the form
\begin{equation}
\overline \psi \vec D\cdot \vec\gamma \psi
\end{equation}
are invariant under this change of variables since $\gamma_0$
anticommutes with $\vec\gamma$.  However, more complicated terms, such
as spatial Wilson hoppings, are generally not invariant under this
change.  This would be a modification of higher order in the lattice
spacing for Wilson fermions.  These are lattice artifacts, presumably
irrelevant to the continuum limit.  Thus, it is only a convention
whether the chemical potiential is inserted as a matrix valued mass as
in Eq.~(\ref{equ:mvm}) or as a direction dependent link term as in
Eq.~(\ref{equ:onlinks}).

If we consider the action as a generalized matrix connecting
fermionic variables
\begin{equation}
S=\overline\psi M \psi,
\end{equation}
the matrix $M$ is not symmetric.  The upper components propagate
forward in time, and the lower components backward.  Even though our
Hamiltonian was Hermitian, the matrix appearing in the corresponding
action is not.  With further interactions, such as gauge field
effects, the intermediate fermion contributions to a general path
integral are generally not positive, or even real.  Of course the
final partition function, being a trace of a positive definite
operator, is positive.  However, depending on the order of operations,
there can be negative intermediate results.  With Monte Carlo methods,
this can lead to uncontrollable fluctuations.  This is the primary
unsolved problem in lattice gauge theory.  With no chemical potential
term, symmetry between particles and antiparticles results in a real
fermion determinant, which in turn is positive for an even number of
flavors.

For our simple Hamiltonian, this discussion has been exact.  The
discretization of time adds no approximations since we could do the
normal ordering by hand.  In general with spatial hopping or more
complex interactions, the normal ordering can produce extra terms
going as $O(1/N^2)$.  In this case exact results require a limit of a
large number of time slices.

In summary, my goals in this discussion were twofold.  First a careful
transfer matrix treatment makes the Wilson projection operator a
natural approach to eliminate temporal doubling.  Second, I have shown
explicitly how to insert a chemical potential term in a path integral.
It is a convention whether this term appears appears as non-symmetric
temporal hoppings or as a matrix-valued effective mass.

\bigskip
{\noindent \bf Acknowledgement:} This manuscript has been authored
under contract number DE-AC02-98CH10886 with the U.S.~Department of
Energy.  Accordingly, the U.S. Government retains a non-exclusive,
royalty-free license to publish or reproduce the published form of
this contribution, or allow others to do so, for U.S.~Government
purposes.

\end{document}